\documentstyle[multicol,epsf,aps]{revtex}
\begin{document}

\title{
Dynamical Properties in the Bilayer Quantum Hall Ferromagnet
}\par

\author{
Tatsuya Nakajima$^{\ast}$
}\par

\address{
Department of Physics, University of Texas at Austin, Austin,
Texas 78712
}\par

\date{February 23, 2001}

\maketitle

\begin{abstract}

The spectral functions of the pseudospin correlation functions in 
the bilayer quantum Hall system at $\nu=1$
are investigated numerically,
where the pseudospin describes the layer degrees of freedom.
In the pseudospin-ferromagnetic phase,
the lowest-energy excitation branch 
is closely connected with the ground state
through the fluctuations of pseudospin $S_y$ and $S_z$,
and it plays a significant role 
on the tunneling properties in this system.
For the system with very small tunneling amplitude
and layer separation smaller than the critical one,
the system-size dependence of 
calculated spectral function $A_{y z}$
suggests the superfluidity on the tunneling current
in the absence of impurities. 

\end{abstract}

\begin{multicols}{2}
\narrowtext


Recently the bilayer quantum Hall (QH) system~\cite{review} 
has often been investigated 
from the viewpoint of quantum ferromagnetism.
In fact,
the pseudospin can be introduced to this system
by a simple definition of assigning the upper/lower layers
to the pseudospin $\uparrow /\downarrow$~\cite{boson,moon}, 
and the pseudospin ferromagnetism can be realized for
the total Landau level 
filling $\nu = 1/m$ ($m$: odd integers).
In particular, the $\nu=1$ bilayer QH system has extensively been
studied both experimentally and theoretically~\cite{review}.

The pseudospin rotational symmetry in this system is degraded
because of the difference between the intralayer and interlayer
Coulomb interaction for a finite layer separation $d$ 
(and of the interlayer tunneling).
Since the two layers tend to have
equal numbers of electrons to reduce the interaction energy,
the $z$ component of the total pseudospin
(half the difference in the numbers)
tends to vanish, $\langle S_z \rangle = 0$,
in the ground state.
Thus the pseudospin will lie in the $xy$ plane
in the pseudospin space, where
the system maintains an invariance under rotations about the $z$
axis, {\it i.e.}, the SU(2) symmetry
is reduced to U(1)~\cite{csdl}.

The electron correlation alone pushes the ground state for
$\nu = 1/m$ 
towards a ferromagnetic one~\cite{mmm},
so the bilayer $\nu =1/m$ QH system behaves like an easy-plane 
itinerant-electron ferromagnet.
The U(1) symmetry is further degraded
in the presence of the interlayer tunneling,
since the tunneling amplitude
behaves like a magnetic field acting on the pseudospin,
as seen in the tunneling Hamiltonian,
$H_{\rm T}=-\Delta _{\rm SAS}\,S_x$.
Here $\Delta _{\rm SAS}$ is the energy difference
between the symmetric and antisymmetric combinations
of isolated layer states and is proportional to 
the interlayer-tunneling amplitude~\cite{boson}.

So a bilayer QH system is characterized
by two dimensionless parameters,
$d/l$ and $\Delta _{\rm SAS}/(e^2/\epsilon l)$
(\,$l$: the magnetic length, 
$\epsilon$: dielectric constant of the host material\,).
For $\nu = 1$,
the phase diagram against these two parameters
is obtained experimentally~\cite{murphy},
and the diagram shows
that the QH state disappears for $d > d_c$ and that
the critical separation $d_c$ increases
as $\Delta _{\rm SAS}$ increases.

Within this QH effect region, 
the ground state for $\nu=1$ 
evolves continuously from
tunneling-dominated (single-particle like)
to correlation-dominated (many-body like)
as $\Delta _{\rm SAS}$ is decreased,
and there is no intervening non-ferromagnetic
region between the two regimes~\cite{na1}.
Moreover, within this pseudospin-ferromagnetic region,
the ground-state wavefunction is analytically given
as a spin-squeezed state~\cite{sss},
where the fluctuation $\Delta S_z$
is supressed by the electron correlation.

By using a sample with very weak interlayer 
tunneling ($\Delta _{\rm SAS}/(e^2/\epsilon l) \sim 10^{-6}$),
Spielman {\it et al} \,have recently showed
the huge resonant enhancement
of the zero bias tunneling conductance~\cite{sp1}.
The possibility of the excitonic superfluidity was also
discussed because of the tunneling spectroscopy
reminiscent of the Josephson effect,
of which controvertial predictions have been done~\cite{csdl}.
Very recently, 
further tunneling-spectroscopy measurements 
were performed with the in-plane magnetic field 
applied to the system,
and the observation of a linearly-dispersing Goldstone mode
in a bilayer QH ferromagnet was claimed~\cite{sp2}.
This in-plane field effect has been discussed in 
some theoretical papers~\cite{stern2,balents,fogler}.

In case of very weak interlayer tunneling as realized
in these experiments~\cite{sp1,sp2},
the U(1) symmetry of the system is expected to recover
almost completely.
Some simple model systems with this symmetry 
have been investigated and the tunneling properties
have been discussed~\cite{stern2,balents,fogler}.
The Chern-Simons Landau-Ginzburg approaches 
for bilayer QH systems~\cite{bone,demler,kim} were 
also performed for the case of vanishing interlayer tunneling.

On the other hand,
some discussions based a BCS-like wavefunction 
with a phase variable $\phi$, which determines the orientation of
the pseudospin magnetic moment, have also been done~\cite{moon}.
This wavefunction, however, is compatible with the SU(2) symmetry,
and does not reflect the U(1) symmetry of the system
as well as the Hartree-Fock approximation does not.
Thus an approach that fully reflects the system's symmetry
and takes small (but non-zero) tunneling amplitudes
into consideration has yet to come
for the study of the tunneling-current properties in this system.
That is exactly the purpose of the present paper.

By the exact diagonalization method 
for small-size systems,
we obtain the spectral functions
of pseudospin correlation functions numerically.
We show that in the pseudospin-ferromagnetic phase 
the lowest-energy excitation branch 
is closely connected with the ground state
through the fluctuations of pseudospin $S_y$ and $S_z$,
and that it plays a significant role for the tunneling properties
in this system. 
For the system with very small tunneling amplitude
and layer separation smaller than the critical one,
the system-size dependence of 
calculated spectral function $A_{y z}$
suggests the superfluidity on the tunneling current
in the absence of impurities.


Our numerical calculations were done 
for bilayer spherical systems~\cite{mmm},
where each surface of the spheres 
is passed through by $2S$ flux quanta.
For the case of $\nu=1$ we consider in this paper,
$2S=N-1$ for $N$-electron systems.
In terms of annihilation and creation operators
for the $m$-th spatial orbit in the lowest Landau level
with pseudospin $\sigma$,
the total pseudospin operators are given as
$S_z = \sum _m (
a^{\dagger}_{m \uparrow} a_{m \uparrow}
- a^{\dagger}_{m \downarrow} a_{m \downarrow}
)/2$,
$S_x = \sum _m (
a^{\dagger}_{m \uparrow} a_{m \downarrow}
+ a^{\dagger}_{m \downarrow} a_{m \uparrow}
)/2$,
$S_y = \sum _m (
a^{\dagger}_{m \uparrow} a_{m \downarrow}
- a^{\dagger}_{m \downarrow} a_{m \uparrow}
)/2 i$. 
Pseudospin $S_x$ and $S_z$ are related with
the interlayer-tunneling and interlayer-voltage part of
the Hamiltonian, respectively, as
$H_{\rm T} = - \,\Delta _{\rm SAS} \,S_x$ and 
$H_{\rm V} = e V S_z$.
Pseudospin $S_y$ is related with the interlayer current $I$
as $I \propto \Delta _{\rm SAS} \,S_y$.
Our calculations were performed for the cases of 
zero interlayer voltage ($V=0$).

The spin density operators are given as
$S_z({\bf \Omega})=[\Psi _{\uparrow}^{\dagger} ({\bf \Omega}) 
\Psi _{\uparrow} ({\bf \Omega})
-\Psi _{\downarrow}^{\dagger} ({\bf \Omega}) 
\Psi _{\downarrow} ({\bf \Omega})]/2$,
$S_x({\bf \Omega})=[\Psi _{\uparrow}^{\dagger} ({\bf \Omega}) 
\Psi _{\downarrow} ({\bf \Omega})
+\Psi _{\downarrow}^{\dagger} ({\bf \Omega}) 
\Psi _{\uparrow} ({\bf \Omega})]/2$,
$S_y({\bf \Omega})=[\Psi _{\uparrow}^{\dagger} ({\bf \Omega}) 
\Psi _{\downarrow} ({\bf \Omega})
-\Psi _{\downarrow}^{\dagger} ({\bf \Omega}) 
\Psi _{\uparrow} ({\bf \Omega})]/2i$,
where 
a unit vector ${\bf \Omega}$ denotes a position on each sphere. 
Their `Fourier transforms' are given by
$S^{\alpha}_{L M} = \int d {\bf \Omega} \,
Y_{L M} ({\bf \Omega}) \,S_{\alpha} ({\bf \Omega})$,
where $\alpha = x,\,y,\,z$, and   
$Y_{L M} ({\bf \Omega})$ is a spherical harmonics
for angular momentum $L$ and its $z$-component $M$.
In particular, the components
for $L=M=0$ are related with the pseudospin operators as
$S^{\alpha}_{0 0} = S_{\alpha}/\sqrt{4 \pi}$.

The spectral functions~\cite{he,yogesh} 
of pseudospin correlation functions 
at zero temperature are given as
\begin{eqnarray}
A_{\alpha \beta} \,(L, \omega) &=& \sum _i
\langle \Phi _0 |\left (
S^{\alpha}_{L M}
\right ) ^{\dagger}
| \Phi _i \rangle
\langle \Phi _i |
S^{\beta}_{L M}
| \Phi _0 \rangle \nonumber \\
& & \quad \quad \quad \quad \times \delta
(\omega - E_i + E_0) ,
\label{eqn:spcf}
\end{eqnarray}
where $| \Phi _0 \rangle$ and $| \Phi _i \rangle$ are
the ground state and the $i$-th eigenstate of
the Hamiltonian, respectively.
The diagonal element $A_{\alpha \alpha} (L, \omega)$ 
among these spectral functions is real and non-negative
because of their definitions in Eq.~(\ref{eqn:spcf}).
Among the off-diagonal elements,
$A_{y z} (L, \omega)$ is closely related with
the finite-wavevector conductance $g$ 
on the interlayer current
because $g(L, \omega) \propto i^{-1}
\,\Delta _{\rm SAS} \,A_{y z} (L, \omega)$
from the Kubo formula.
We note that $A_{y z} (L, \omega)$ is purely imaginary
and that the spectral functions in Eq.~(\ref{eqn:spcf}) 
are independent of $M$ 
because of the rotational symmetry of the system.


Table~\ref{tab1} shows the fluctuations of pseudospin $S_z$ and 
expectation values of pseudospin $S_x$ in the $\nu = 1$ ground state 
for $N=10$, $d/l=1.0$, and 
$t \equiv \Delta _{\rm SAS}/(e^2/\epsilon l)=10^{-4}$, 
$10^{-5}$, $10^{-6}$.
For such small $\Delta _{\rm SAS}$,
the values are proportional to $\Delta _{\rm SAS}$,
because $\Delta _{\rm SAS}$ is proportional 
to the interlayer tunneling amplitude
and behaves like a magnetic field along the $x$ axis 
in the pseudospin space.
Table~\ref{tab1} indicates that the U(1) symmetry 
in the easy-plane ferromagnet is recovered almost completely 
in case of very small $\Delta _{\rm SAS}$.
In fact, for $t \lesssim 10^{-3}$, 
intra- and inter-layer correlations between electrons
in the ground state 
are found to be little dependent on the value of $t$. 
On the other hand,  
the interlayer-tunneling features 
are strongly dependent on it.

\begin{figure}[t]
\begin{center}
\leavevmode\epsfxsize=164mm \epsfbox{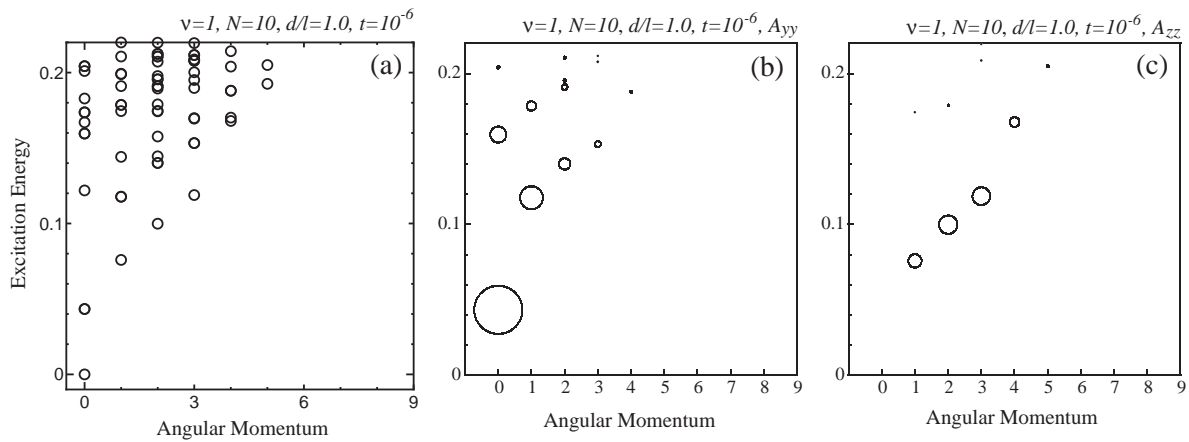}
\end{center}
\vspace*{0.7cm}
\caption{
The excitation spectra and spectral functions
in the $\nu = 1$ bilayer QH system are shown 
for $N=10$, $d/l=1.0$, and  
$t \equiv \Delta _{\rm SAS}/(e^2/\epsilon l) = 10^{-6}$.
(a) The energy levels are shown against the total angular momentum 
by open circles.
The strength for each excitation 
in the spectral functions (b) $A_{yy}\,(L, \omega)$ 
and (c) $A_{zz}\,(L, \omega)$ is shown 
against the angular momentum $L$ and excitation energy $\omega$
by the area of the corresponding circle.
The energy is measured from the ground-state energy
and the unit is $e^2/\epsilon l$.
}
\label{fig1}
\end{figure}

In Fig.~\ref{fig1}, the excitation spectra and spectral functions
in the $\nu=1$ bilayer QH system are shown 
for $N=10$, $d/l=1.0$, and $t=10^{-6}$.
The excitation spectra are shown by open circles 
in Fig.~\ref{fig1}(a).
In Fig.~\ref{fig1}(b) and (c),
the strength for each excitation 
in the spectral functions $A_{yy}\,(L, \omega)$ 
and $A_{zz}\,(L, \omega)$ 
is shown against the angular momentum $L$ 
and excitation energy $\omega$
by the area of the corresponding circle, respectively.
We note that the energy is measured from the ground-state energy
and that the unit is $e^2/\epsilon l$.

The lowest-energy excitation branch
can be seen for $0 \le L \le 3$ in Fig.~\ref{fig1}(a).
We made sure that the energy values
in this branch (in particular, the one
for $L=0$) are quite dependent on the system size $N$ 
and that these ones decrease as $N$ increases.
This branch is expected to be seen as
a linearly-dispersing excitation mode~\cite{disp}
for sufficiently large $N$
and very small $\Delta _{\rm SAS}$.
This pseudospin-wave excitation branch
is closely connected with the ground state 
through the fluctuations of pseudospin $S_y$ and $S_z$.

As seen in Table~\ref{tab1},
the U(1) symmetry is recovered almost completely 
for $t \lesssim 10^{-4}$. 
In fact, we made sure that 
the result for the spectral function $A_{x x}$ 
is almost the same as that for $A_{y y}$ shown in Fig.~\ref{fig1}(b), 
and that each pair of excited states corresponding to 
$A_{x x}$ and $A_{y y}$, respectively, is nearly degenerate.
On the other hand, as seen in Fig.~\ref{fig1}(c),
most of the spectral weights for $A_{z z}$ are occupied by the 
lowest-energy excitation branch.
This is also the case for $A_{y z}$,
because $A_{y z}$ and $A_{z z}$ 
include the same factor for each term in Eq.~(\ref{eqn:spcf}).

In Table~\ref{tab2},
the strength of $i^{-1} A_{y z} \,(L, \omega)$
in the lowest-energy 
excitation branch in the $\nu=1$ pseudospin ferromagnet
is shown for $N=10$, $d/l = 1.0$, and 
$t = 10^{-4}$, $10^{-5}$, $10^{-6}$.
It indicates that  
these values of strength are proportional to $t$, {\it i.e.}, to 
$\Delta _{\rm SAS}$.
Thus the tunneling conductance is considered to
be proportional to the square of $\Delta _{\rm SAS}$,
because $g(L, \omega) \propto i^{-1} \Delta _{\rm SAS} 
A_{y z} \,(L, \omega)$.
It is also found that the strength increases 
as angular momentum $L$ decreases.
That is, 
the long-wavelength part of the pseudospin-wave excitation mode
plays a significant role on the interlayer-tunneling properties
in this system.

In Table~\ref{tab3},
the strength of $i^{-1} A_{y z} \,(L, \omega)/t$
in the lowest-energy excitation branch 
in the $\nu = 1$ pseudospin ferromagnet
is shown for $d/l = 1.0$, $L=0$, $1$, $2$, and $N=6$, $8$, $10$.
It has a system-size dependence varying approximately as
$N^{\lambda}$ ($\lambda \simeq 3$).
This is also the case for $d/l = 0.5$.
For larger layer separations such as $d/l=1.5$ and $2.0$,
however, such a system-size dependence
is not obtained.
In fact, for $d/l=1.5$ and $2.0$,
the strength increases {\it or decreases} \,as $N$ increases,
and the system-size dependence is much weaker than that for 
$d/l=0.5$ and $1.0$.

For sufficiently small $\Delta _{\rm SAS}$ 
(\,$\lesssim 0.01 e^2/\epsilon l$\,),
it is known that the pseudospin ferromagnetism is destroyed 
for layer separation $d$ larger than 
the critical one $d_c \simeq 1.3 l$~\cite{boson,john}.
Among the values of layer separation adopted by us,
$d/l=0.5$ and $1.0$ belong to the case of $d < d_c$, and 
$d/l=1.5$ and $2.0$ satisfy $d > d_c$.
As mentioned above,
our results for $d < d_c$ show a system-size dependence 
of the spectral function $A_{y z}$ varying approximately 
as $N^{\lambda}$ ($\lambda \simeq 3$),
which is much stronger than usual $N$-linear dependence.
On the other hand,
the possible superfluidity on the tunneling current in this system 
is considered to show an $N$-square dependence.
It is not clear whether 
this discrepancy between our result and predicted one results from
small system sizes in our calculations or not.
At least, however,
the system-size dependence of $A_{y z}$
in our results is much stronger than $N$-linear one
and so suggests the superfluidity on the tunneling current
in the absence of impurities.
For clarifying interlayer tunneling properties in this system,
further studies of the impurity effects~\cite{stern2}
and numerical calculations for larger systems are desired.

For $d > d_c$,
such a strong system-size dependence does not appear 
in our numerical results.
Significant differences between the two cases of 
$d > d_c$ and $d < d_c$ are expected to become clearer
in larger systems
because of the differences in the system-size dependence.
These qualitative differences between the two cases 
are consistent with 
the results in a recent experiment~\cite{sp1}.
Lastly we note that 
the spectral weights of $i^{-1} A_{y z}$ 
occupied by the lowest-energy excitation branch
decrease rapidly
as the layer separation increases.



The author thanks Allan H. MacDonald, John Schliemann, 
and Yogesh Joglekar for fruitful discussion with them.
He is also grateful to Hisatoshi Yokoyama 
for valuable conversations on spectral functions,
and to Anju Sawada
for valuable experimental information about
the bilayer QH system.
He was supported for the research in the USA by
Japan Society for the Promotion of Science.

\begin{table}
\begin{tabular}{||c|c|c|c||}
& $t = 10^{-4}$ & $10^{-5}$ & $10^{-6}$ \\
\hline
$\Delta S_z$ & $6.845 \times 10^{-3}$ &
$6.845 \times 10^{-4}$ & $6.845 \times 10^{-5}$ \\
\hline
$\langle S_x \rangle$ & $4.165 \times 10^{-2}$ &
$4.165 \times 10^{-3}$ & $4.165 \times 10^{-4}$ \\
\end{tabular}
\caption
{
The fluctuations of pseudospin $S_z$
and expectation values of pseudospin $S_x$ 
in the $\nu = 1$ ground state 
are shown for $N=10$, $d/l = 1.0$, and
$t \equiv \Delta _{\rm SAS}/(e^2/\epsilon l) = 10^{-4}$,
$10^{-5}$, $10^{-6}$.
For such small $\Delta _{\rm SAS}$,
the values are proportional to $\Delta _{\rm SAS}$.
}
\label{tab1}
\end{table}

\begin{table}[b]
\begin{tabular}{||c|c|c|c||}
& $t = 10^{-4}$ & $10^{-5}$ & $10^{-6}$ \\
\hline
$L=0$ & $16.02 \times 10^{-4}$ & 
$16.02 \times 10^{-5}$ & $16.02 \times 10^{-6}$ \\
\hline
$1$ & $10.56 \times 10^{-4}$ & 
$10.56 \times 10^{-5}$ & $10.56 \times 10^{-6}$ \\
\hline
$2$ & $8.328 \times 10^{-4}$ & 
$8.329 \times 10^{-5}$ & $8.329 \times 10^{-6}$ \\
\hline
$3$ & $4.806 \times 10^{-4}$ &
$4.806 \times 10^{-5}$ & $4.806 \times 10^{-6}$ \\
\end{tabular}
\caption{
The strength of the spectral function $i^{-1} A_{y z} \,(L, \omega)$
in the lowest-energy excitation branch 
in the $\nu = 1$ pseudospin ferromagnet
is shown for $N=10$, $d/l = 1.0$, and 
$t \equiv \Delta _{\rm SAS}/(e^2/\epsilon l) = 10^{-4}$,
$10^{-5}$, $10^{-6}$.
The strength increases as angular momentum $L$ decreases, and
it is proportional to $t$. 
Most of the spectral weights are occupied 
by this pseudospin-wave excitation branch.
}
\label{tab2}
\end{table}

\begin{table}
\begin{tabular}{||c|c|c|c||}
& $N=6$ & $8$ & $10$ \\ 
\hline
$L=0$ & 4.389 & 8.670 & 16.02 \\
\hline
$1$ & 2.144 & 5.225 & 10.56 \\
\hline
$2$ & 1.617 & 4.150 & 8.329 \\
\end{tabular}
\caption{
The strength of $i^{-1} A_{y z} \,(L, \omega)/t$
in the lowest-energy excitation branch 
in the $\nu = 1$ pseudospin ferromagnet
is shown for $d/l = 1.0$, $L=0$, $1$, $2$, and $N=6$, $8$, $10$,
where $t \equiv \Delta _{\rm SAS}/(e^2/\epsilon l)$.
It has a system-size dependence varying approximately as
$N^{\lambda}$ ($\lambda \simeq 3$).
}
\label{tab3}
\end{table}

\end{multicols}
\widetext

\end{document}